\documentclass[epjST]{svjour}

\usepackage{graphics}

\usepackage[numbers]{natbib}

\begin{document}
\title{Interests Diffusion on a Semantic Multiplex}
\subtitle{Comparing Computer Science and American Physical Society communities}
\author{Gregorio D'Agostino\inst{1}\fnmsep\thanks{\email{gregorio.dagostino@enea.it}} \and Antonio De Nicola\inst{1} }
\institute{ENEA - CR "Casaccia"}

\abstract{
Exploiting the information about members of a Social Network (SN) represents one of the most attractive
and dwelling subjects for both academic and applied scientists. The community of Complexity Science and especially those researchers working on multiplex social systems are devoting increasing efforts to outline general laws, models, and theories, to the purpose of predicting emergent phenomena in SN's (e.g. success of a product).  On the other side
the semantic web community aims at engineering a new generation of advanced services tailored to specific people needs. This implies defining constructs,
 models and methods for handling the semantic layer of SNs.  We combined models and techniques from both
the former fields to provide a hybrid approach to understand a basic (yet complex) phenomenon: the propagation of
individual interests along the social networks. Since information may move along different social networks, one should take into account a multiplex structure. Therefore we introduced  the notion of "Semantic Multiplex". 
In this paper we analyse two different semantic social networks represented by authors publishing in the Computer Science and those in the American Physical Society Journals. 
The comparison allows to outline common and specific features.} 
\maketitle
\section{Introduction}
\label{intro}

\begin{figure}
\begin{center}
\resizebox{0.80\columnwidth}{!}{  \includegraphics{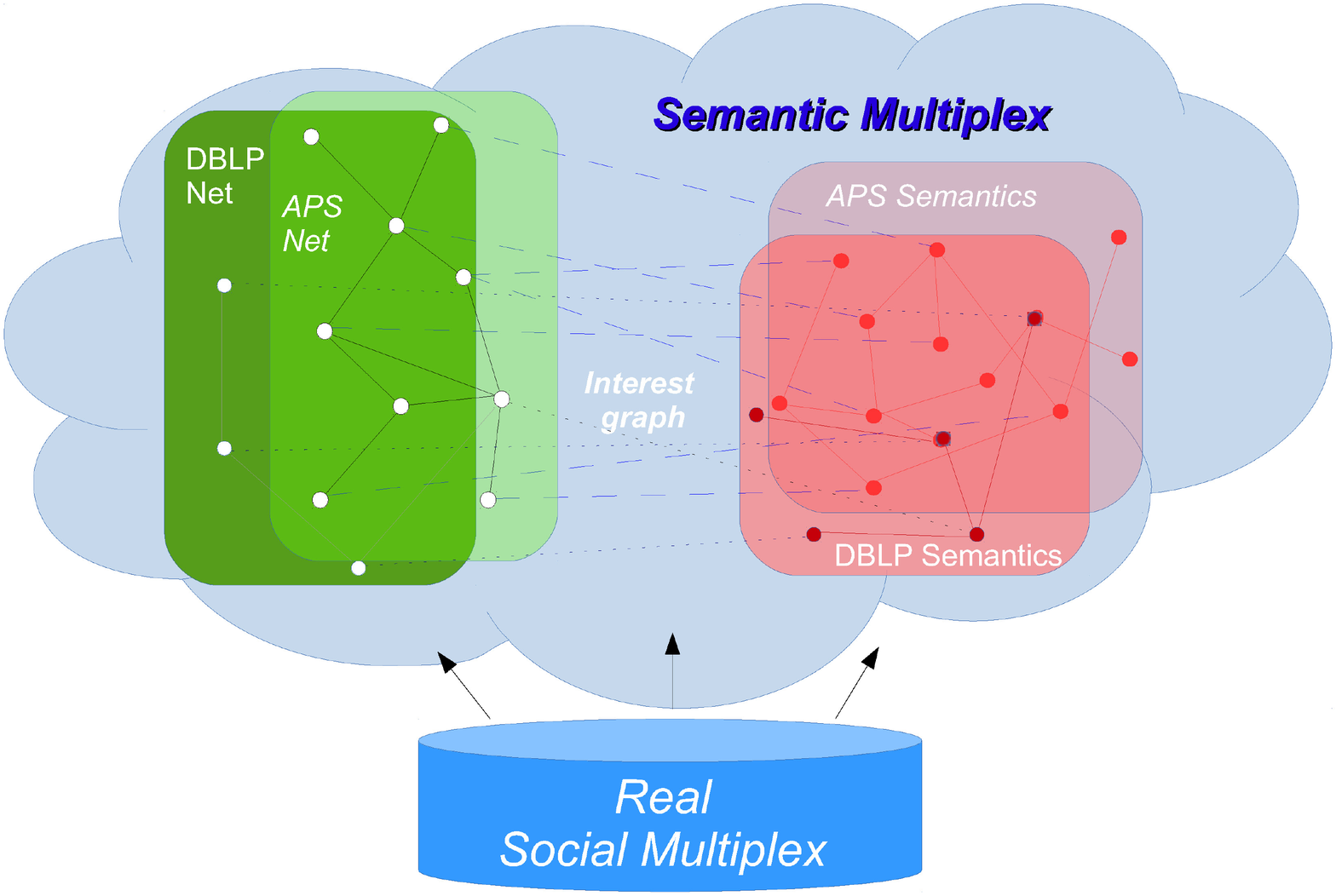} }
\caption{Diagrammatic representation of the "Semantic Multiplex" studied in the present work. Authors belong to one or both of the Computer Science and APS community. They do exhibit scientific social links inside their community and between the two communities; the domain of interest of Computer Science and Physics do also exhibit internal and external semantic links; finally authors interests are represented by links between them and concepts in the 
integrated domain of interest.}
\end{center}
\label{fig:GeneralMultiplex}   
\end{figure}

Understanding how and to what extent people influence (and are influenced by) the different social networks they belong to, it is a very important issue. Profiling people may allow improvement in public services, while representing a powerful means to enhance marketing capabilities by vendors. In the present paper, interest represents a general concept that may refer to very detailed entities such as real product on the market or to abstract entities such as music or literature. Some authors refer to "memes" as basic ideas propagating on social networks, however the definition of interest employed here is more general. 

In this paper we combine semantic  and complexity techniques to provide insights on the interest diffusion on social networks. Our framework consists of the following main components: a method to gather information about the members; methods to perform semantic analysis of the domain of interest; a procedure to infer members' interests; and a model for interests propagation in the network. 

We studied interest diffusion on single domain Semantic Social Networks in some previous works 
\cite{dagostino2015, dagostino2015C}. 
This work represents a first attempt to provide insights on a Semantic Multiplex. We are introducing the former term to refer to a representation of a social network with multiple relationship channels (i.e. a multiplex) linked to a semantic representation of a domain of interest. We studied the special case of the American Physical Society and Computer Science scientific communities, that, in fact, form a "Semantic Multiplex". The links in the scientific social network were limited to co-authorship; members were attributed links to the domain of interest according to their publications; finally links between the concepts were attributed by the semantic analysis of publications. It is worth stressing that both the domain of interest and the social network do exhibit a typical multilayer structure as each member may belong to different social networks;  conversely, each concept may belong to several semantic layers. The interests represent links between members and concepts. Therefore a Semantic Multiplex is a bipartite graph were each of the two halves are multiplexes \cite{Boccaletti20141}. Fig. \ref{fig:GeneralMultiplex} provides a pictorial representation of our conceptualisation instanced for the test case:
members may belong to one of two networks of  authors and they exhibit some links to the semantic space representing the (stratified) domain of interest.  
The two sets, authors and interests, form a bipartite network; while the social and the semantic networks do exhibit internal links representing friendship and closeness of concepts, respectively. This means that three different types of links are accounted. Alternatively one may image a "Semantic Multiplex" as a stacking of different "Semantic Social Networks" as introduced in \cite{dagostino2015}.

Along the line with our previous research, we assumed diffusion to be the basic mechanism driving the interest spreading in the social network. We further assumed that members have a limited tendency to change their interests $x_i$, a capability to influence their neighbours $x_{ji}$ (i.e. authority), and a tendency $x_{si}$ to be influenced by ongoing general trends $L_{s}$:
\begin{equation}
\label{DiscreteInterestPropagation}
L_i (c_k,t+\Delta t) = \left[1-x_i-x_{is}\right] \cdot L_i(c_k,t)+ \frac{1}{|N_i|}\cdot \sum_{j \in N_i} x_{ij} \cdot L_j(c_k,t)+x_{is} \cdot L_s(c_k,t)
\end{equation}
where $L_i=L_i(c_k)$ represents the level of interest of member $i-th$ in a topic $c_k$ at time $t$ and the $x_{ij}$,  $x_{ij}$ represent the average opinion change due to neighbours and general trend, respectively.  When time increment is brought to zero, one is left a set diffusion-like equations.
\begin{equation}
\label{InterestPropagationEquation}
\frac {d}{dt} L_i (c_k,t) = -\left[a_i+a_{is}\right] \cdot L_i(c_k,t)+ \frac{1}{|N_i|}\cdot \sum_{j \in N_i} a_{ij} \cdot L_j(c_k,t)+a_{is} \cdot L_s(c_k,t).
\end{equation}
 \noindent The $a_{ij}$ and  $a_{sj}$ are the rate of interest change in unit time and $a_i=\sum_j a_{ij}$.  The total attitude of a member to change her or his interests in unit time is represented by $a_i+a_{si}$ i.e. the total susceptibility. 

In the case of scientific SN's, one can estimate $L_i(c_k)$'s by the fraction of publications produced in topic $c_k$. This allows to apply the general model to the semantic social network represented by the
authors and the subjects of their publications. Authors form a social network by co-authorship, while the topics of the publications can be represented as a semantic network which can be organised as a taxonomy. The interest of an author in a subject is expressed by a publication in that field.

Applying the above defined paradigm to real set of data, provides an analytic tool to measure individual features, such as members' susceptibilities and authorities. Authority is quantity that measure the capability to influence other members that we quantify by $A_i=\sum_j a_{ji} $. It is worth stressing that $A_i $ and $a_i$ are different because the probability  of being influenced (i.e. susceptibility) is different from the probability to influence other people: $a_{ji} \ne a_{ij}$. The social network endowed by the $a_{ji}$ weights, may be seen a weighted directed network.

\section{The Computer Science  and American Physical Society Research Communities}

Although the approach applies to any type of social network, here we test it against two different scientific research communities: the Computer Science (CS) and the American Physical Society (APS) ones.
More precisely, we employ the DBLP (Digital Bibliography and Library Project) and the APS databases as exclusive certified sources. This means that a semantic representation of a domain of interest and co-authorships networks for the two selected cases were extracted, respectively, from the DBLP\footnote{DBLP: Digital Bibliography \& Library Project. http://www.informatik.uni-trier.de/~ley/db/} computer science bibliography and the APS (American Physical Society) dataset\footnote{APS  datasets can be retrieved at http://journals.aps.org/datasets}. Both the sources above provide (real-time) rich datasets covering a long time period on both the social network and authors interests. Such amount of data guarantees that, if existing, the interest diffusion phenomenon can be observed and the individual features estimated. Despite some limitations related to the simplicity of the diffusion theory and to some intrinsic limits in the semantic analysis, the basic mechanism for interest contagion is a diffusion process \cite{dagostino2015}. 
 
Several scientific papers using the DBLP dataset demonstrate its validity as a standalone case study (\cite{Elmacioglu:2005}, \cite{Staudt:2012}, \cite{Han:2010}, \cite{kudvelka2012social}, and \cite{amblard2011temporal}). However most of them address structural and topological properties of the social network and only partially the social network dynamics. The second standalone case study concerns basically the research social network in the field of physics.  

The analysis of the APS publications is a also representative application of the general interest diffusion phenomenon as it covers a large time period (even larger than that covered by DBLP). Furthermore, as the DBLP case study, the scientific community deems the APS dataset to be relevant for studies in the field of social and complex networks as demonstrated by the existence of several outstanding papers using it (e.g. \cite{Boccaletti20141}, \cite{deville2014career}). In particular, the following APS journals were considered for this analysis: Physical Review (PR), Physical Review A (PRA), Physical Review B (PRB), Physical Review C (PRC), Physical Review D (PRD), Physical Review E (PRE), PRI, Physical Review Letters (PRL), Physical Review Special Topics - Accelerators and Beams (PRST-AB), Physical Review X (PRX), and Reviews of Modern Physics (RMP).

Here the goal of the experimental evaluation is to test the theory (that is eq. \ref{InterestPropagationEquation}) against the two real cases and comparing the resulting characteristics of the two communities.
In order to perform the analysis, we need to acquire the information about the topics defining the scope of the computer science domain and the evolution dynamics of both the social relationships and the interests of the authors.
In principle, the above information could be extracted from different sources, however the DBLP dataset and APS provide both the information through a single or a set of XML documents [35]. 

The analysis of each community followed our methodology for experimental assessment of social networks and, specifically, consisted of the following steps:
\begin{enumerate}
\item{\textit{papers selection}, according to their type (e.g., journal, conference, book chapter) and year;} 
\item{\textit{interests and topics identification};}
\item{\textit{papers indexing};}
\item{\textit{identification of social network topology and its temporal evolution};}
\item{\textit{semantic profiling of the members}};
\item{\textit{analysis of the trends};}
\item{\textit{assessment of the validity of the interests propagation theory and estimation of individual features}.}
\end{enumerate}

\subsection{Papers Selection} 
\label{sec:treatable}

The DPLP and APS databases are evolving entities. Results presented in this work refer to the datasets as published by November 2013. For DBLP the observation period was limited to the years from 1950 to 2012. In such a temporal range, the number of considered papers is 2246098 and the authors 1337195. In order to study the evolution of authors' interests it is necessary to observe some change in their semantic profile during time; therefore only authors that published papers in, at least, two different years can be analysed. Those authors were named "treatable". It is worth noting that only 519886 authors out of 1337195 are treatable as far as 2012. It is reasonable to image that this is mainly due to students that just publish one work and then leave the world of research for other activities. It should be noted that not-treatable authors are intrinsically untreatable, i.e. they are so independently from the specific capability of a suitable set of topics to index papers.  

Similarly to DBLP, the APS (American Physical Society) dataset provides a list of scientific papers in journals of physics. The results presented here refer to the dataset provided by the American Physical Society including papers up through 2013. The observation period was limited to the years from 1955 to 2005. In this temporal range, the number of considered papers is 357553. 
The member identification in the DBLP suffers from ambiguities. There are authors' names under which papers by different members are gathered (polysemy) and, viceversa, there are authors that sign different papers with slightly different names (synonymy).  This issue is particularly relevant for Asian names. 
The problem is currently approached by different authors with promising results \cite{ferreira2012brief, wang2014unified, milojevic2013accuracy, schulz2014exploiting, petersen2014inequality}.

A suitable wider set of sources will allow to detect a more complete set of topics and, hence, to define more accurate profiles. This problem is mostly solved for the APS  dataset since it provides information on affiliations that can be used to disambiguate authors, improve the quality of the analysis and, hence, of the overall assessment of the theory.

The members of the social network identified with a disambiguation engine (see \cite{dagostino2015C}) are 257391. 
Only 133058 members out of 257391 are treatable  in the considered time period. 
%


\subsection{Identification of Interests and Topics}
\label{sec:IdInTop}

In principle the text of all communications form the potential corpus  supporting the semantic network. However we limited the analysis to the titles. There are other possible choices such as including the abstract or the introduction. Nevertheless it is worth noting that a lot of information contained in the papers (and, specifically, in those sections) do not refer to their specific contents, but to the general state of the art in the field and, hence, the semantic analysis of the full text (or the abstract and the introduction) could include spurious terms not specific to the subject.  Finally the analysis of millions of full papers is extremely time consuming and it may be not sustainable. For all the reasons above, limiting the analysis to the titles seems the most appropriate and handy choice.

A TermExtractor web application \cite{sclano2007termextractor} and a software platform developed for natural language processing \cite{dagostino2015C} were employed to extract preliminary multi-lexemes from the corpus of titles. A lexeme is basically a set of words sharing a common root, or, more precisely, the class of equivalence of a word and all its linguistic inflected forms. This tool is based on natural language processing techniques \cite{Navigli:2004:LDO:1105710.1105712}
and allows extracting the shared terminology of a community from the available documents in a given domain.
Then the multi-lexemes were validated by human processing devoted to remove general-purpose ones that are not specific of the computer science (or APS) domain and to merge synonyms. 
The list of candidate topics was also double-checked to assure lack of possible polysemic multi-lexemes.
The resulting list of lexemes is the best approximation of the set of basic topics  $\{c_k\}$ we were able to build up.
Recently, based on Latent Dirichlet Allocation, new methods have been employed  for automated topic extraction \cite{lancichinetti2014high}.
These novel techniques will possibly improve the quality of the set.

The selection of the basic set of topics $C=\{c_1, c_2, \ldots, c_N\}$ plays a crucial role to "tame" the domain of interest. 
It is worth stating that, in order the diffusion theory to work, the $c's$ must form a "basis" for the algebra of interests. The most relevant relations among concepts (and, hence, among interests) are generalization (specialization) and similarity. From the algebraic point of view these represent inclusion relationships.  The two constraints here imposed on the set $C$ are "completeness" and "independence" respectively. 

A set of concepts will be named "complete" when each concept can be seen as the union of a subset of basic concepts:
\begin{equation}
\label{decomposition}
\forall c \exists \{ i_1, i_2, \ldots, i_m\} : c=\cup_{k=1}^m c_{i_k}.
\end{equation}
\noindent this means that each interest is the combination of a set of the basic interests. 

On the other side a set of concepts will be named "independent" when each pair is disjoint, that is, it does not exist a concept representing a common specialisation of both:

\begin{equation}
\label{independency}
\forall c_i, c_j  : \ c_{i} \cap c_j =\emptyset.
\end{equation}
\noindent When eq. (\ref{independency}) holds, the decomposition of eq. (\ref{decomposition}) is unique.

The basic topics are identified with a subset of the multi-lexemes. The quality of the results strongly depends on the capability of the selected set of multi-lexemes to fullfil the required constraints.

Once the set of topics is assessed, it is possible to attribute a subset of them to each scientific product. Conversely, each topic can be given a frequency as the number of papers referring to it ($\nu (c_k,t)$).

7632 topics were identified for DBLP and 30967 topics for APS.

\subsection{Identification of Social Networks Topology and Semantic Profiling}
\label{sec:sempro}

For each year, the nodes of the social network are given by the authors that have written papers by that year and the edges are given by the co-authorships. According to this assumption, the social network grows with the time. At a given a year, a link is attributed to all authors that have published a product together by that time. Links are not removed and, therefore, the social network shows increasing complexity. It is worth noting that in real life authors stop to publish (e.g., for retirement or job change) or give up their collaboration. However such phenomena was not taken into account since the work was limited to the information available in the DBLP or APS datasets. Moreover, all members are treated on the same ground regardless of their notoriety or scientific production. The former approximations may affect the quality of these results.  

\begin{figure}
\resizebox{0.48\columnwidth}{!}{
\includegraphics{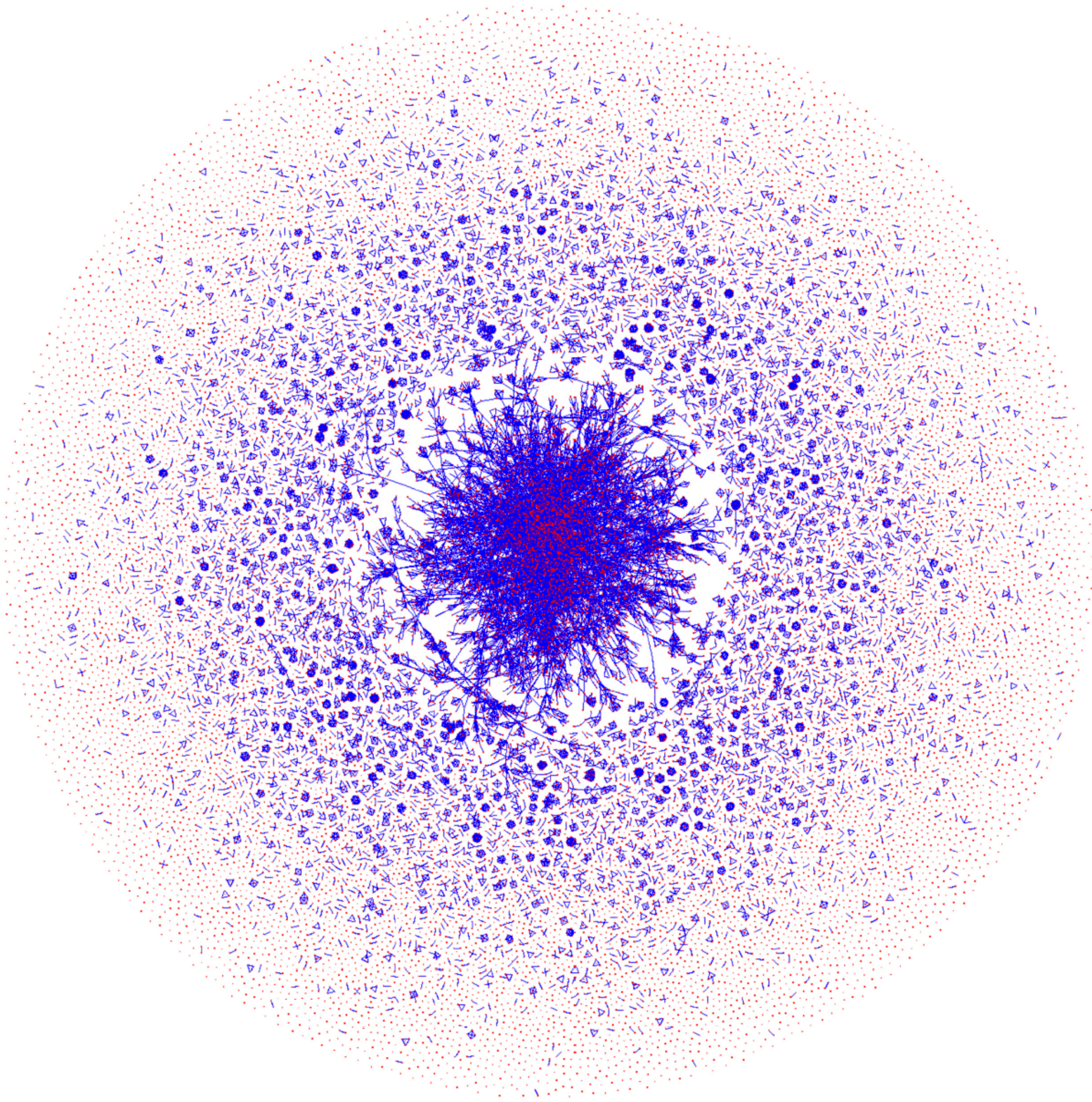}}
\resizebox{0.48\columnwidth}{!}{
\includegraphics{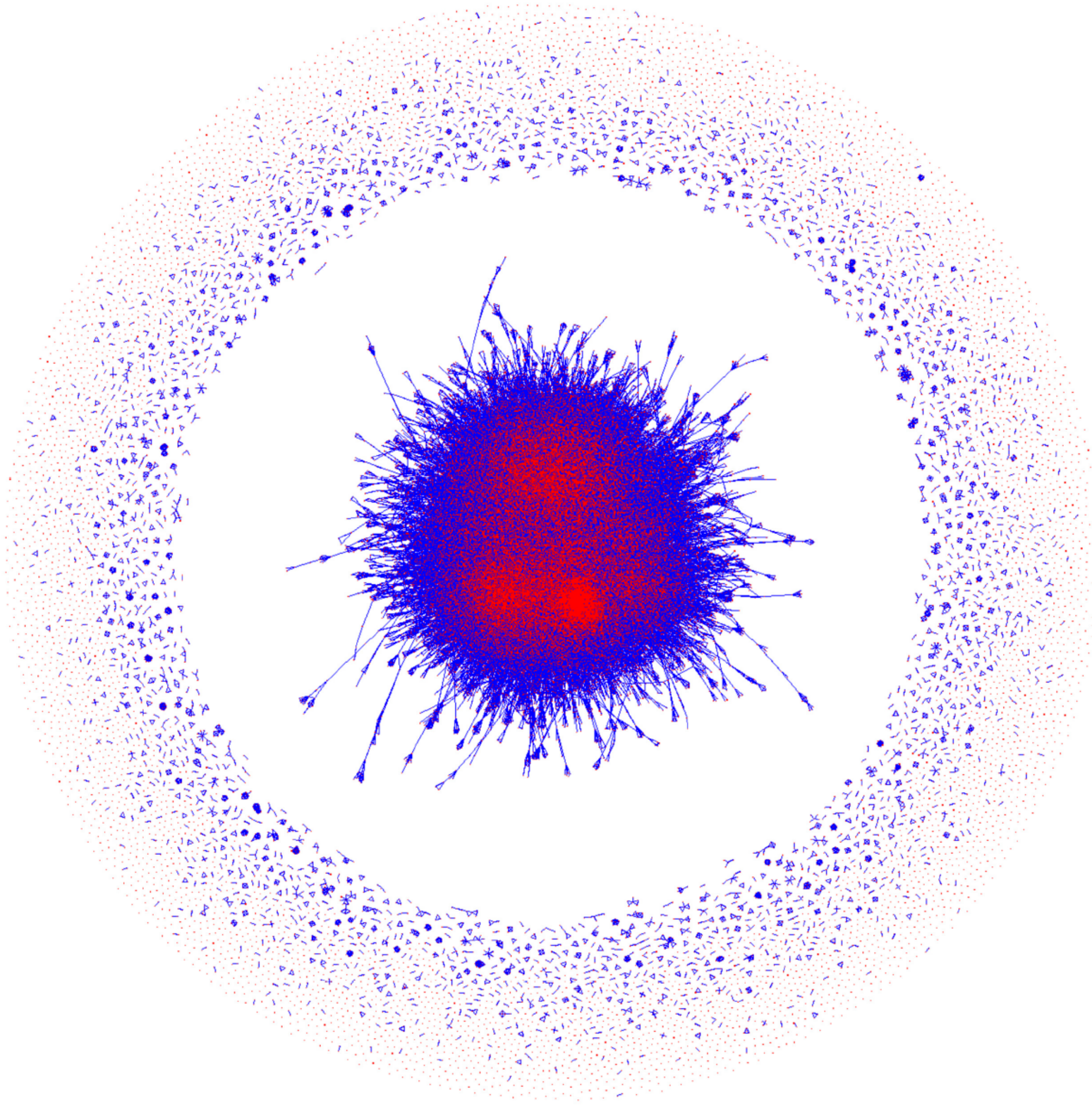}}
\caption{A graph representation of  Computer Science (left) and APS (right) networks: authors are red dots while links are blue. 
Three regions are clearly identified: a huge core, disconnected small communities and a crown of isolated authors.}
\label{fig:SNresults}
\end{figure}

Figure \ref{fig:SNresults} shows snapshots of the structure of the DBLP and APS social networks in 1980. In accordance with what observed in \cite{Kumar:2006}, both the networks can be partitioned in three regions: single authors who do not participate in the network; isolated communities which display star structures; and a huge component characterised by a well-connected core region.

For each year in the observation period, a semantic profile can be given to each author by means of the relative frequencies of "expressions of interest" that in this contest coincide with publications.
Treatable authors are named "semantically treatable" once they acquire a not-trivial semantic profile. Their number is affected by the present capability to identify interests related to the domain and to index all papers. In principle, it is possible to estimate both trends and neighbours susceptibilities of semantically treatable authors.  However, the higher is the number of publications of an author, the more precise is the estimation of susceptibility. By sewing together all the semantic profiles of the different members one achieves the total interest graph.

To reconstruct the time evolution of the social network, for each year, a link to all pairs of members sharing a paper published by that year was attributed. Weights were not attributed to links depending neither on the age of the shared publications nor on their number or scientific relevance. The former are two strong hypotheses that may be relaxed in future (ongoing) work. In fact, human contacts that took place in a remote past may have ceased; moreover, a successful publication may stimulate further scientific common activity more than a coarse one. 

\subsection{Numerical Results}
\label{sec:numresDBLP}

We performed a maximum likelihood fit to extract information on members susceptibility to both trend and neighbours \cite{dagostino2015, dagostino2015C}.
For both the networks, the hypothesis that neighbours influence the future propagation is necessary as it exhibits the maximum likelihood even upon normalising  $\chi^2$ with the number of degrees of freedom. 
The analytical $\chi^2$ minimization  provides best fit quantities for the majority of members, however in some cases those solutions are not feasible (i.e. the [0,1] range constraint is violated). In those cases the
most likely values are achieved at boundaries. Negative and super-unitary values may represent real social characteristics of members (not accounted in the present model) or may result from the incompleteness of the semantic analysis. As it is widely discussed in \cite{dagostino2015}, probably both interpretations contribute. 

\begin{figure}[!h]
\begin{center}
\resizebox{0.49\columnwidth}{!}{
\includegraphics{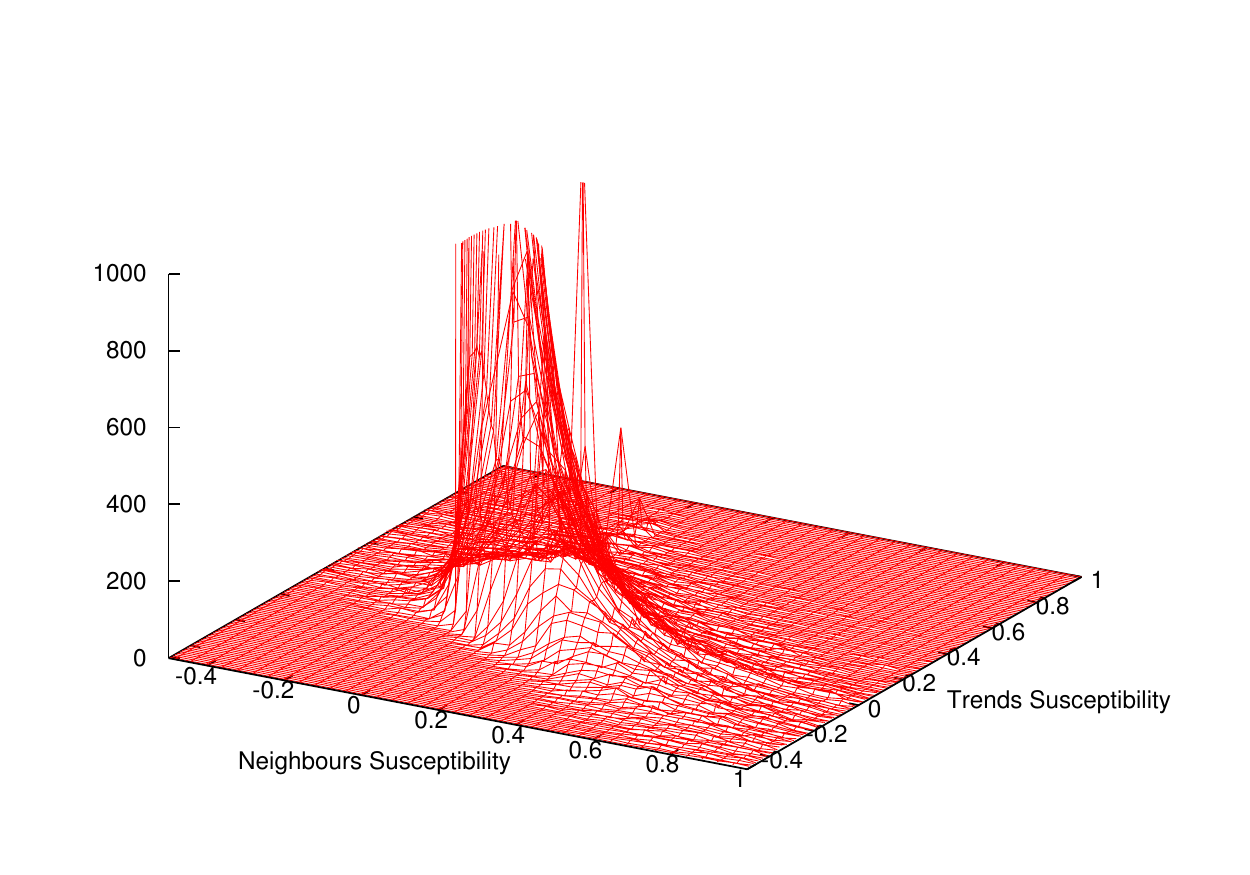}}
\resizebox{0.49\columnwidth}{!}{
\includegraphics{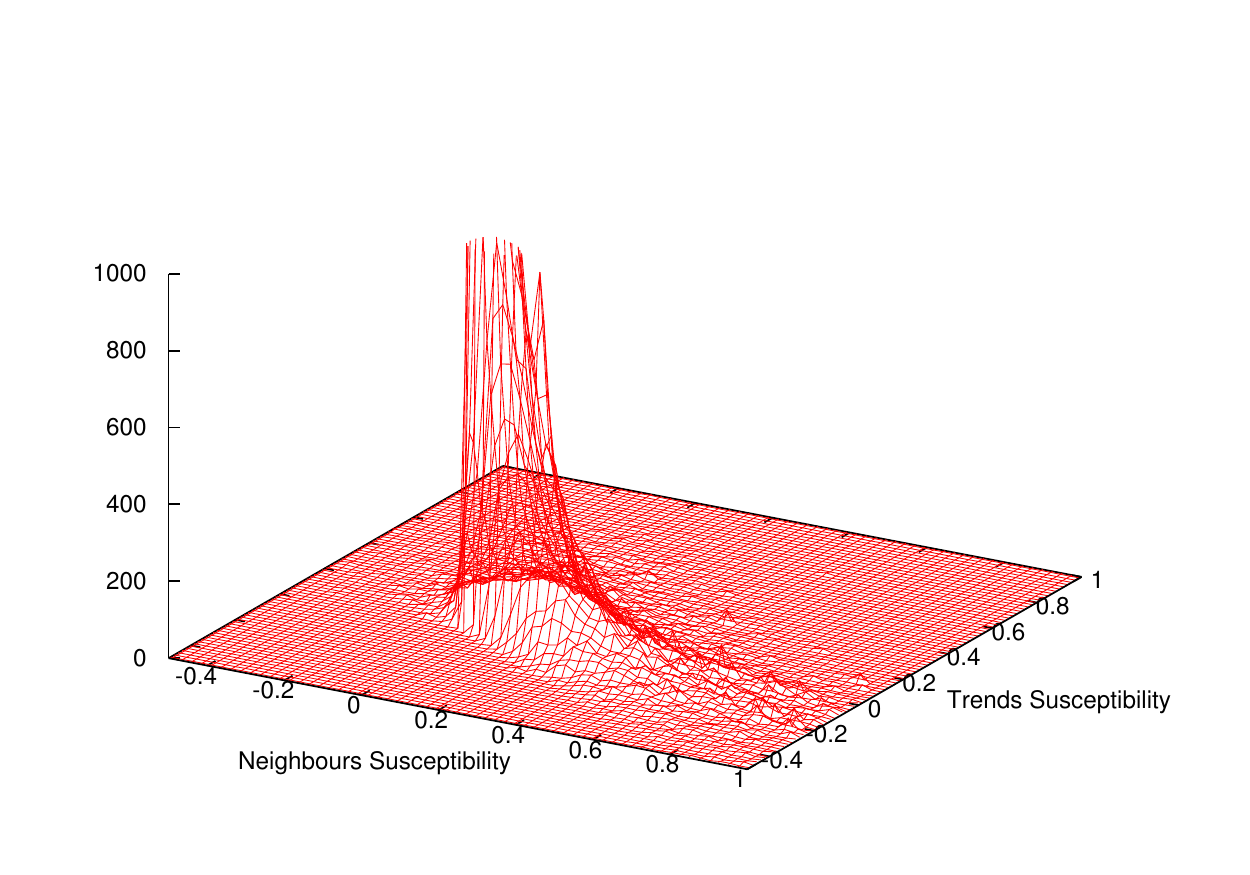}}
\caption{Three dimensional histogram of the frequencies of the fitted $\hat{x}_i$ and $\hat{x}_{is}$ for the computer science case study (left) and for the APS (right) dataset.}
\end{center}
\label{histograms3D}
\end{figure}

The authority coefficients (authorities, shortly) of DBLP authors spans the [0,52] range; their mean value is $\bar a = 0.44$ and its standard deviation is about twice that value (0.89).  
While there certainly exist real authors with some hundred collaborators, some of the observed ones may be fictitious.
It is known that there exist different authors with the same name (given name and family name); 
those people are very often treated as a single author in several datasets. 
This problem is known as "ambiguity" of the papers indexing; it results in gathering different authors into a single member of our social network. 
Due to the many to one transliteration of the original names in latin characters, Asian members are mostly prone to such effect.
The problem is known to affect DBLP data analysis \cite{Han:2010:MKD:1807167.1807333}. Currently, DBLP administrators only use the surname and given name for the disambiguation, while providing room for explicit requests from the authors. Evidently this is not enough to prevent disambiguation.
In order to check this phenomenon, a list of frequent Chinese full names by combining 50 very common Chinese given names \cite{wiki:chinesenames} with 100 frequent Chinese family names \cite{wiki:chinesesurnames} was built.
Several high values of authority correspond to entries associated with the constructed set of frequent Chinese full names. 

Table \ref{FamousAuthors} presents some individual features of some famous authors in computer science. As expected they all exhibit high levels of authority.

Table \ref{APSNobelPrize} reports the neighbours' and trends' susceptibilities and the authority of famous scientists winning the Nobel Prize in Physics. This table shows a relationship between the number of coauthors and the authority even for the most famous scientists. This suggests to consider a different index of authority obtained by normalizing the actual authority parameter against the number of coauthors. This will be further investigated in the future.

\begin{table}[ht]
\caption{Famous authors in Computer Science}
\label{FamousAuthors}
\centering
\begin{tabular}{|c|c|c|c|}\hline
\textbf{Name} & \textbf{$\hat{x}_i$} & \textbf{$\hat{x}_{is}$} & \textbf{Authority $a_i$}\\
\hline
\hline
\textbf{Wil M. P. van der Aalst} & $+0.111$ & $+0.058$ & $+12.809$ \\
\hline
\textbf{Jack Dongarra} & $-0.019$ & $+0.028$ & $+10.259$ \\
\hline
\textbf{John Mylopoulos} & $+0.021$ & $+0.037$ & $+8.852$ \\
\hline
\textbf{Georg Gottlob} & $+0.055$ & $+0.009$ & $+5.081$ \\
\hline
\textbf{Ian Horrocks} & $+0.198$ & $-0.080$ & $+4.835$ \\
\hline
\textbf{Maurizio Lenzerini} & $+0.106$ & $-0.065$ & $+3.128$ \\
\hline
\textbf{Erol Gelenbe} & $+0.184$ & $+0.015$ & $+3.123$ \\
\hline
\end{tabular}
\end{table}

\begin{table}[ht]
\caption{Some authors winning the Nobel Prize in Physics. The table shows, respectively, the name, the year of the Nobel Prize, the neighbours' and trends' susceptibilities, the authority, the number of papers considered for this analysis and the number of considered coauthors.}
\label{APSNobelPrize}
\centering
\begin{tabular}{|c|c|c|c|c|c|c|}\hline
\textbf{Name} & \textbf{Year} & \textbf{$\hat{x}_i$} & \textbf{$\hat{x}_{is}$} & \textbf{$a_i$} & \textbf{Papers} & \textbf{Coauthors}\\
\hline
\hline
\textbf{Steven Weinberg}  & $1979$ & $+0.003$ &  $+0.391$ &  $+0.883$ &  $128$ &  $38$ \\
\hline
\textbf{Carlo Rubbia} & $1984$ & $-0.009$ & $+0.099$ & $31.843$ & $40$ & $289$ \\
\hline
\textbf{David Gross} & $2004$ & $+0.164$ & $+0.155$ & $2.300$ & $10$ & $44$ \\
\hline
\end{tabular}
\end{table}

\begin{figure}[!h]
\resizebox{0.48\columnwidth}{!}{
\includegraphics{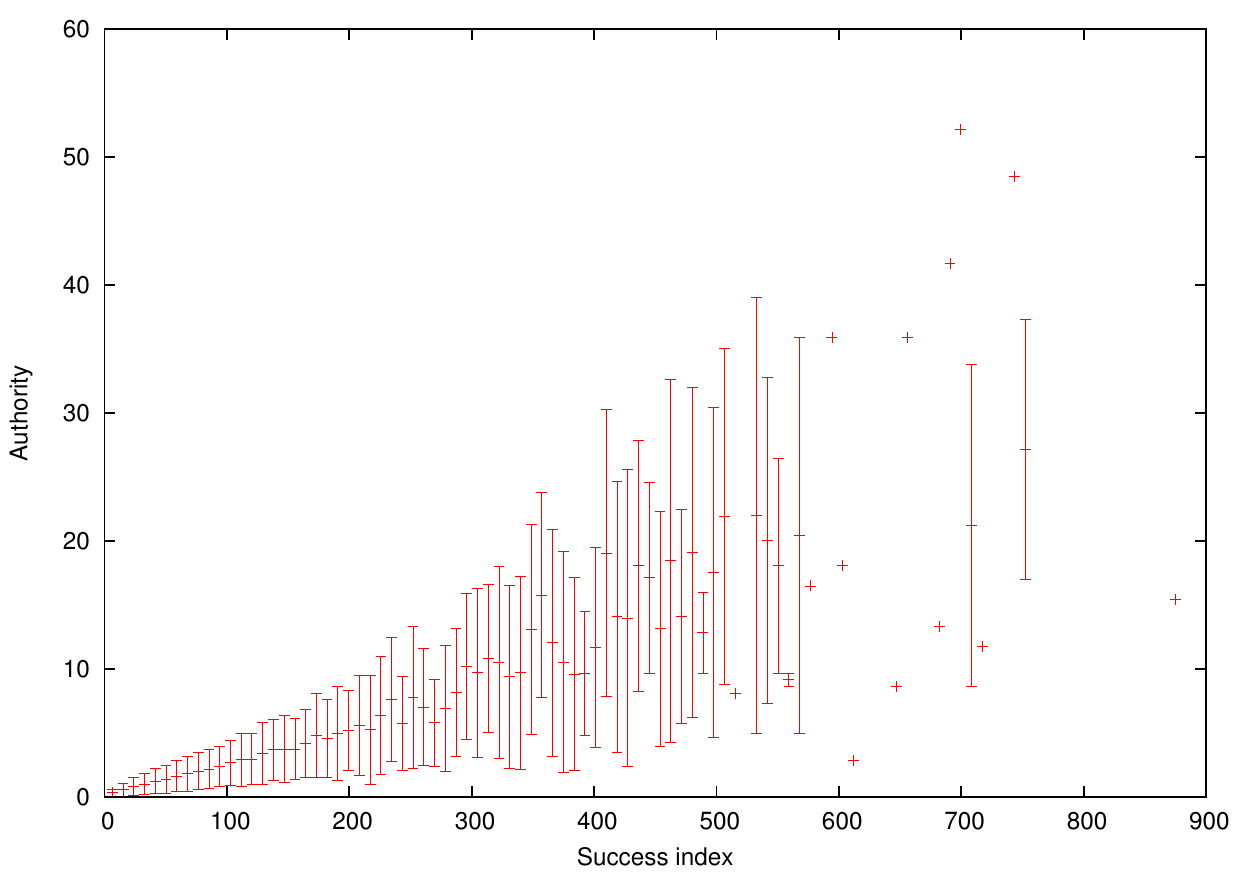}}
\resizebox{0.48\columnwidth}{!}{
\includegraphics{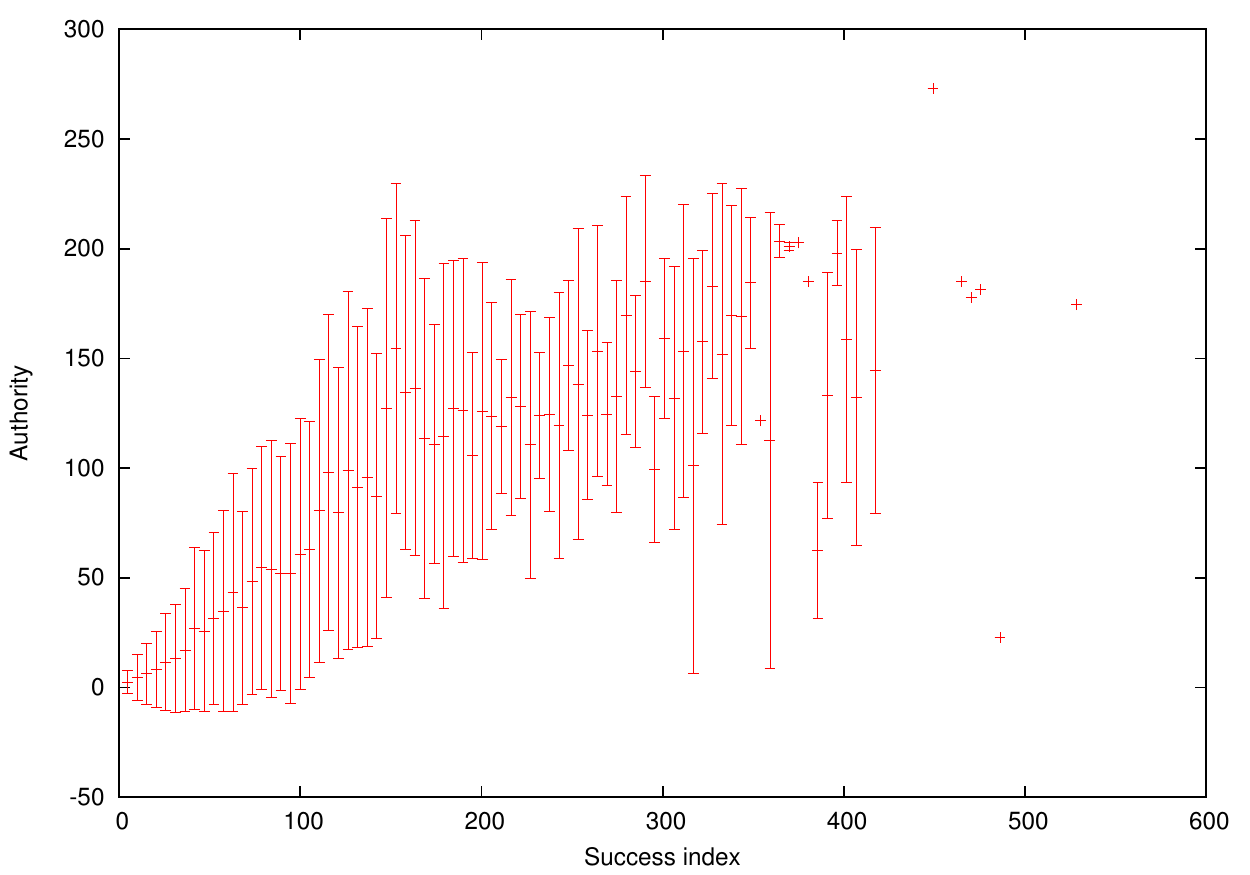}}
\caption{Scatter plots of the authority of different semantically treatable authors versus their success index for the two test cases: computer science (left) and APS (right).}
\label{fig:AuthvsSucc}
\end{figure}

Then the relationship between the success of an author with its authority was investigated. The number of published papers (including proceedings) was employed as an index of success. A more appropriate index should be the total number of citations \cite{Petersen28102014, Wang2013quantifying} or the \textit{h-index} which were not available. As shown in Figure \ref{fig:AuthvsSucc}, the higher is the success index, the higher is the authority. 
We are not able to provide evidence of significant dependence of trende susceptibility upon success. This means that there are successful authors of different types: some of them follow trends; some propose new topics and some continue working mostly on the same topics. 

Beside the limits of the theory there are other factors that need further treatment and possibly hindered the analysis of the DBLP: the semantic analysis and the disambiguation.

Currently, interests are extracted from the titles of the papers by means of natural language techniques. Even assuming that there exists a basic set of disjoint topics covering the domain of interest, the semantic analysis may only lead to an approximation of it. Small sets of basic topics are not capable to index all papers, while larger ones tend to contain synonyms, similar multi-lexemes and, above all, concepts not representing interests (e.g. the fake ones resulting from the words: report, surveys etc). As already discussed, in this experimentation, an effect of the lack of coverage of the domain is the presence of null values of $x_i$ and $x_{is}$. 
This issue affects the quality and the completeness \cite{burton2005semiotic} of the identified topics and has an impact on semantic profile estimation.



\begin{figure}
\begin{center}
\resizebox{0.48\columnwidth}{!}{
\includegraphics{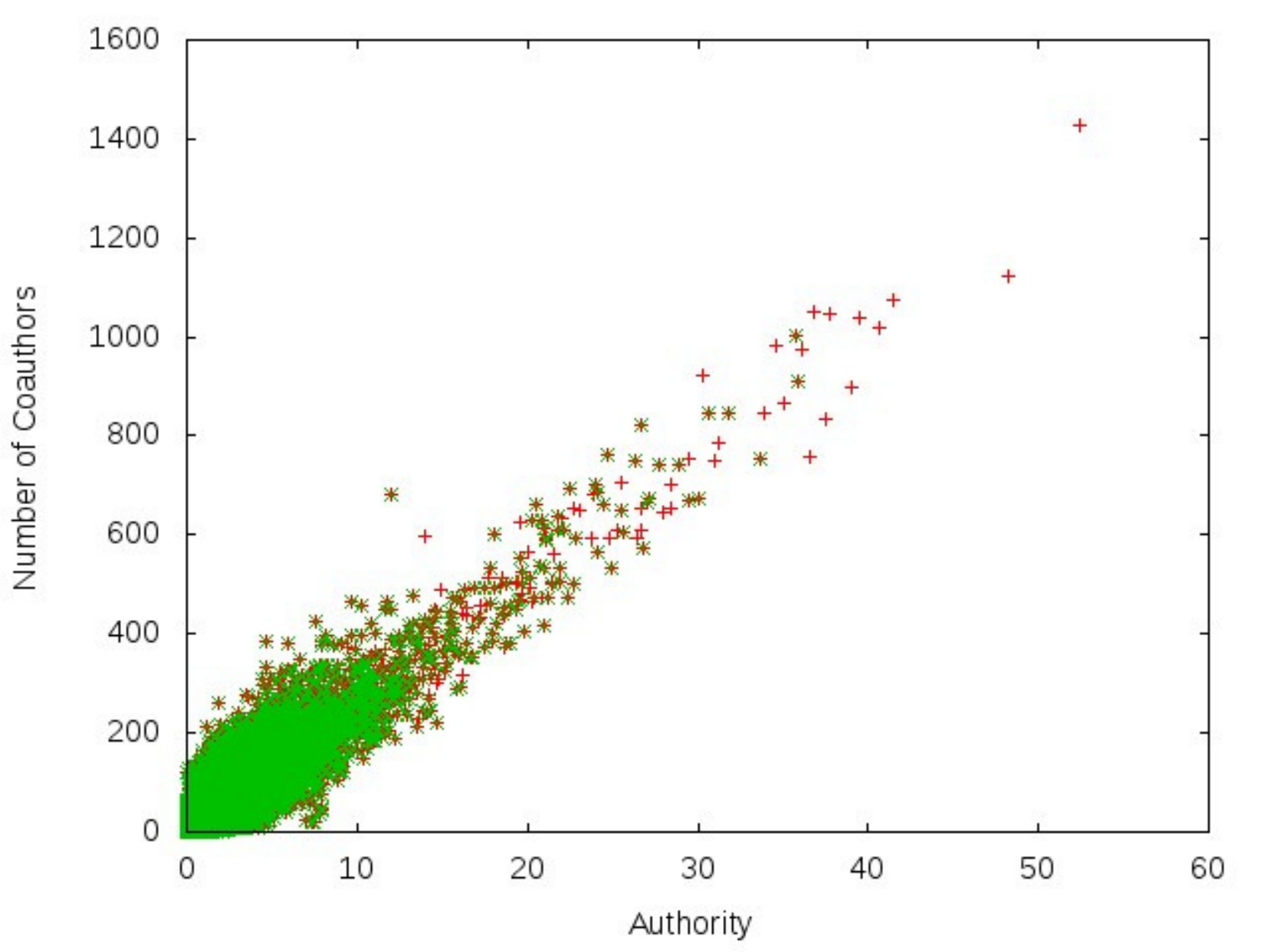}}
\resizebox{0.48\columnwidth}{!}{
\includegraphics{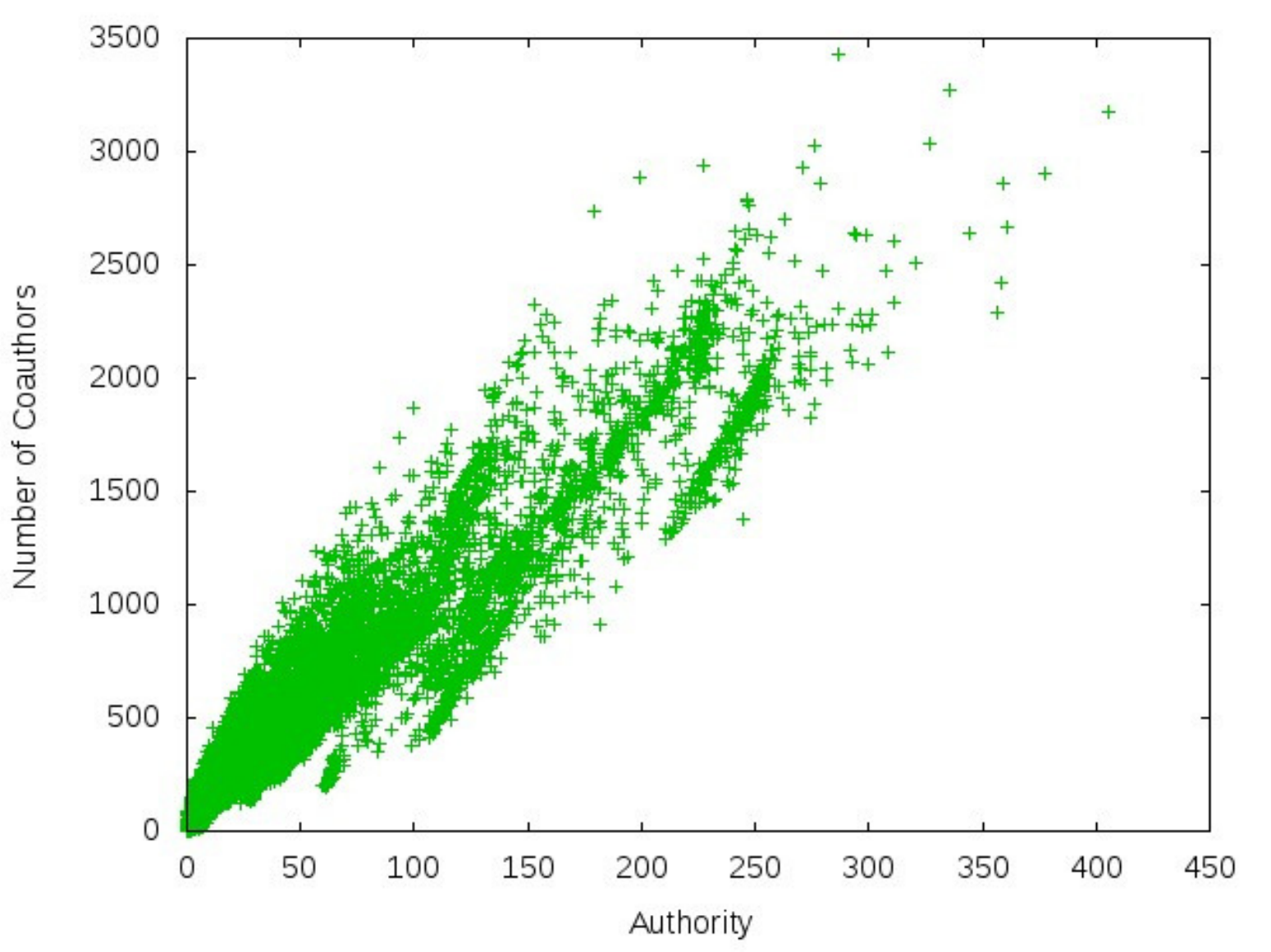}}
\caption{Scatter plot of co-authorship vs authority for both datasets (DBLP left, APS right). Red crosses correspond to authors with very common asian names who where impossible to disambiguate.}
\end{center}
\label{fig:CoauthVSauthority}
\end{figure}

Figure \ref{fig:CoauthVSauthority} presents the scatter plot of the number of coauthors versus the authority for each author in the DBLP and APS datasets. In both cases, authors with high authority have a huge number of coauthors. However, while the disambiguation plays a relevant role for the DBLP case, it is reasonable that in APS dataset, due to availability of the affiliation, authors with large neighbours are real. 
This is probably due to the existence of several papers (especially in high energy physics) describing experiments involving hundreds of scientists.

\section{Comparing the two scientific communities}
\label{sec:comparison}

Despite some slight overlap, the computer science and the APS community are definitely distinct. Generally speaking the authors of both communities exhibit the same average susceptibility to
trends, however the APS authors are twice more susceptible to co-authors influence. This means that the new ideas do circulate more effectively in the APS community. In other words physicists tend to share the new interests more than computer scientists. Possibly due to the large teams involved in very expensive facilities such as those at CERN in Geneve, ERFS in Lausanne, or neutron facility ion Rutherford, real large collaborations are observed in the APS community, while some much smaller group exists in computer science. In both cases authority is related to success in the average even though wide fluctuations are observed. On the other side the trend susceptibility does not seem to be related to the success in any of the two communities. In fact we observe successful authors that have been working for long time on a specific topic and others that have changed their field of interest continuously. 

\begin{figure}
\begin{center}
\resizebox{0.48\columnwidth}{!}{
\includegraphics{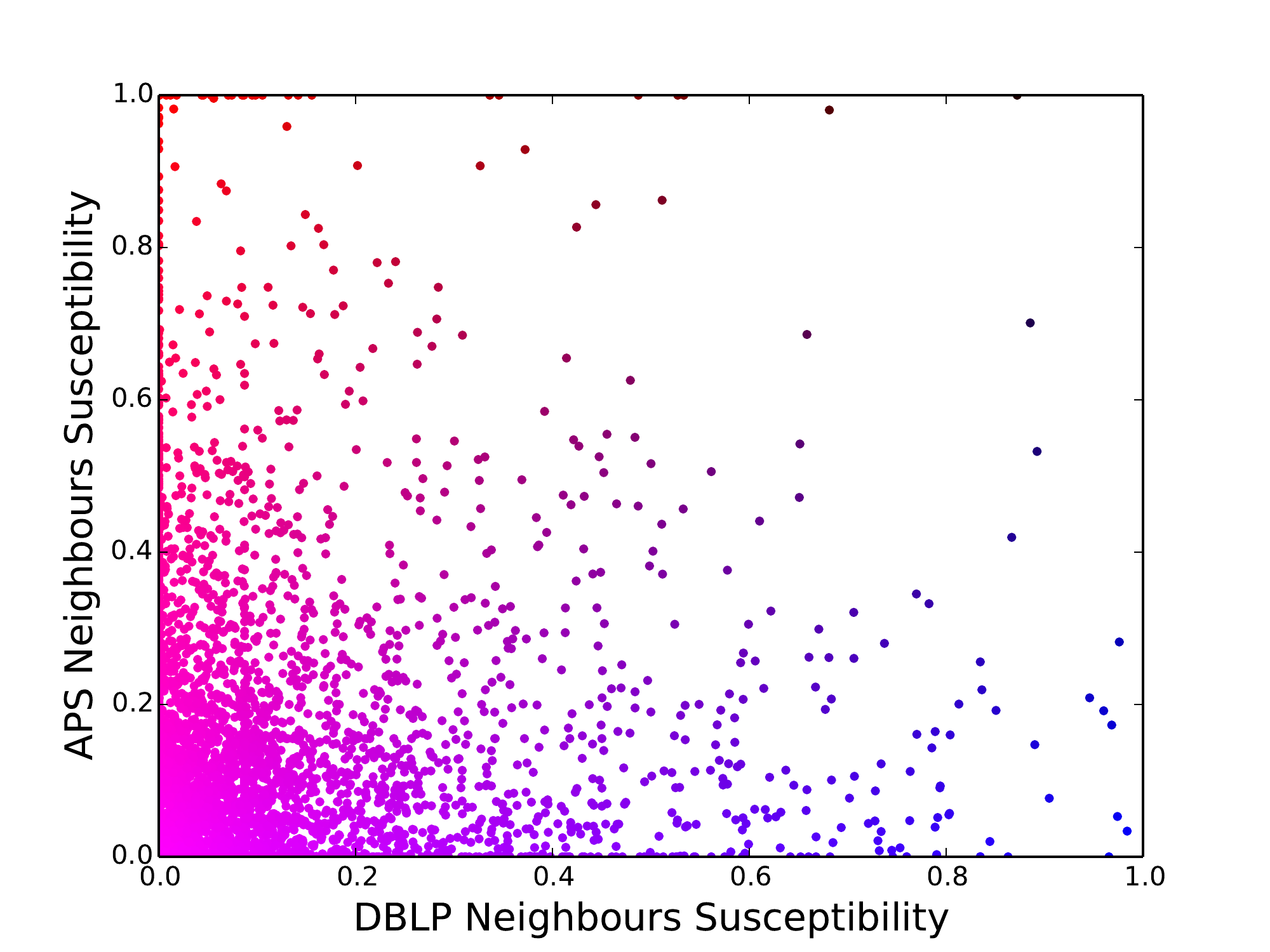}}
\resizebox{0.48\columnwidth}{!}{
\includegraphics{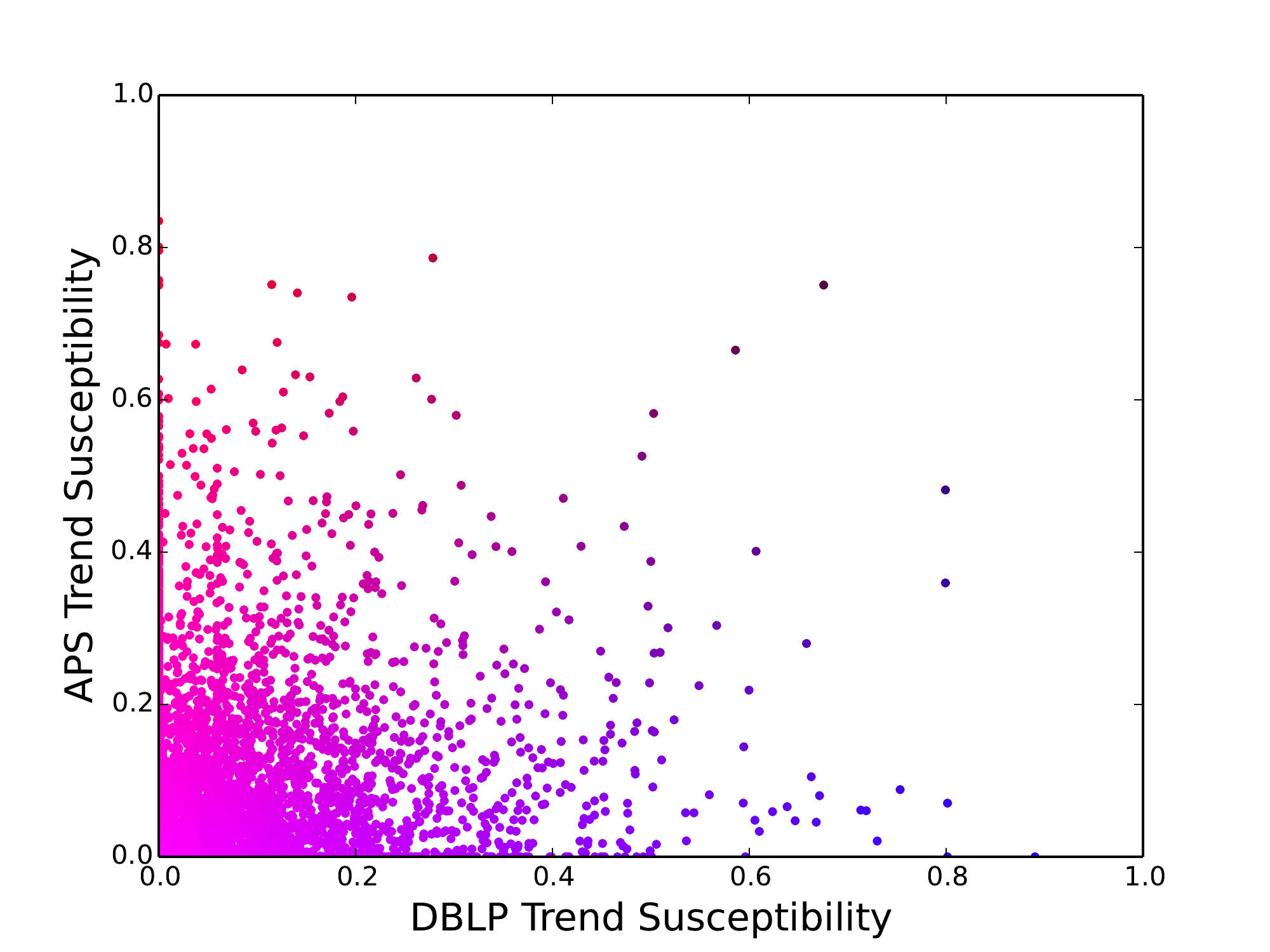}}
\caption{Scatterplot of neighbours' (left) and trends' (right) susceptibilities as measured for one author in the computer science and physics datasets.  
To improve legibility color coding is also used: blue corresponds to authors with high susceptibility in Computer Science, while red indicates high susceptibility on APS activities.}
\end{center}
\label{fig:SuscComparison}
\end{figure}

Table \ref{table:Comparison} reports some average quantities for the set of authors belonging to both communities. As can be seen, the tendency of APS community is confirmed also for this special sample, as well as the basic susceptibility to general trends. Again even the same authors do exhibit a larger average authority, thus indicating that it is the habit of publish more in the  community and not a specific human characteristic responsible for the effect. The correlations between the susceptibilities and the authorities as measured in the different disciplines are all slightly positive. This means that there exists a very soft effect due to human character, but the leading effect is the position occupied in the community to determine the authority of susceptibility. This result is also evident from Figure \ref{fig:SuscComparison} which shows the measured characteristics of the different authors as measured in the APS community versus the same quantity measured in the computer science community. If the measured features were related to some human intrinsic characteristics, one expects the point to concentrate on the diagonal. On the contrary they are concentrated on the axes. Moreover the few points falling on the diagonal  tend to be concentrated at low values, especially for the authorities. This confirms the general idea that authority is strongly dependent on the field and normally the highest in a field the lowest in an other.

\begin{table}[ht]
\caption{\bf{Comparing Member Characteristics of Computation (DBLP) and  APS communities.} Data in parentheses refer to the equivalent value calculated on the whole DBLP or APS sample.}
\label{table:Comparison}
\begin{tabular}{|c|c|c|c|}
\hline
\hline
\textbf{$Community$} & \textbf{Average Neigh. Susc.} & \textbf{Average Trend Susc.} & \textbf{Average Authority} \\
\hline
DBLP & 0.089  (0.093) & 0.070 (0.071)  &  1.079 (0.439) \\
APS & 0.14  (0.163)  & 0.087  (0.062) &  5.75 (6.27)      \\
Correlation & 0.037  & 0.040 &  0.057  \\
\hline
\end{tabular}
\end{table}

\begin{figure}
\begin{center}
\resizebox{0.8\columnwidth}{!}{\includegraphics{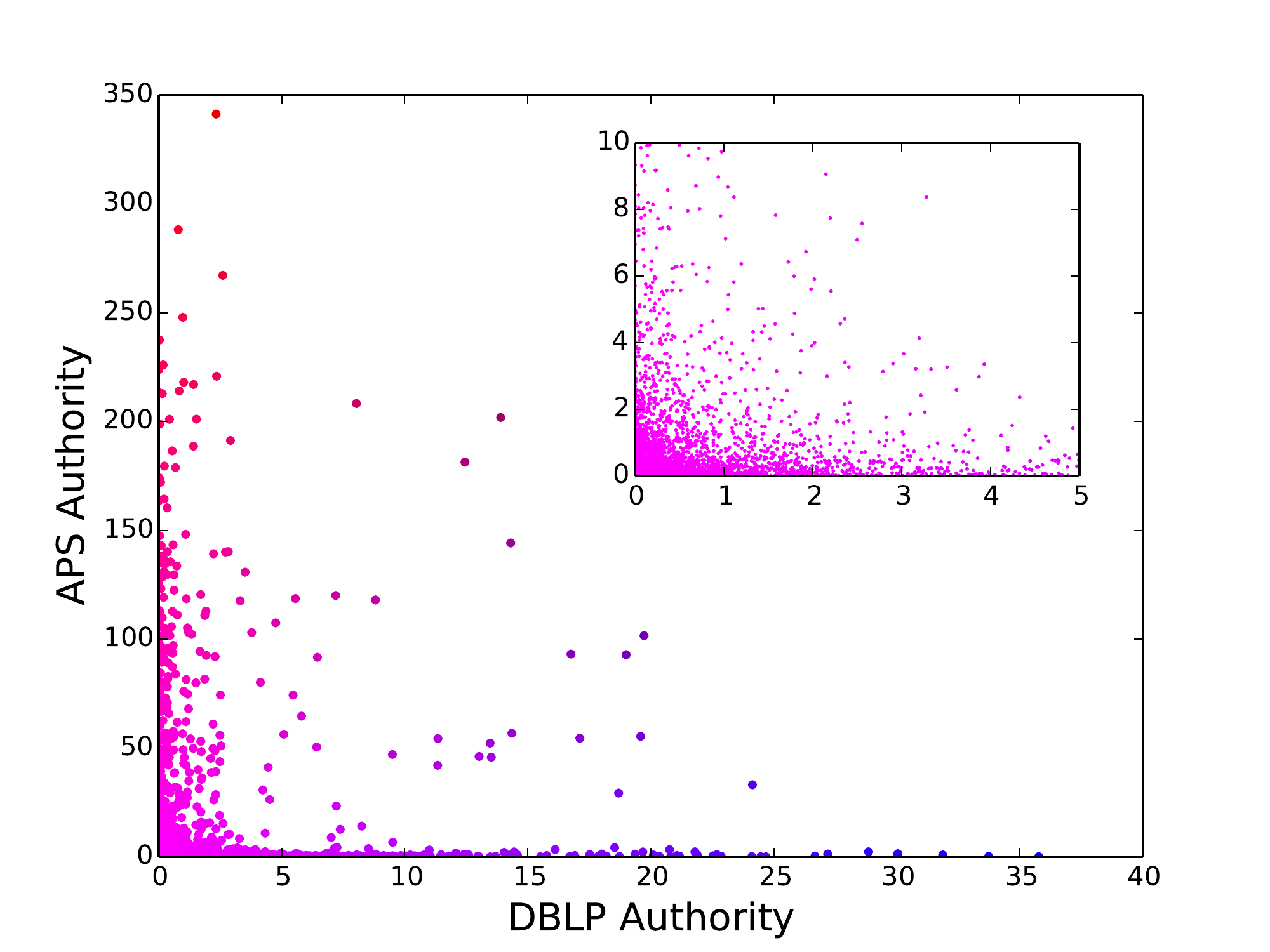}}
\caption{Authorities measured in the two different dataset for the authors writing in both the computer science and physics disciplines. The colours of the different points are
set in a way that blue corresponds to authors with high authority in computer science, while red indicates authority in APS topics. The inset focuses on the most populate region.}
\end{center}
\label{fig:AuthComparison}
\end{figure}

\section{Discussion}
\label{sec:disc}

This work represents a first application of a model for diffusion of interests to a real social network observed from two different layers.
A social network endowed with a semantic domain of interest and its relation with members is named "Semantic Social Networks", since here we are dealing with several interacting  Semantic Social Networks
we introduced the concept of "Semantic Multiplex" (abridged form for Semantic Social Multiplex) which represents a multiplex (a social network with different relation layers) endowed with a semantic representation of a domain of interest.

As a propagation model, we assumed that interests spread according to a diffusion mechanism, while being continuously created. 
The human behaviour is described by means of two basic behavioural characteristics (the susceptibility and the authority)
 that quantify the tendency to be influenced by "friends" and the environment. By that means also the capability to influence others is inferred.

The model was tested against two interdependent scientific co-authorship networks: the APS and computer science communities. This set provides a prototype of a "Semantic Multiplex".

Despite some intrinsic limitations of the model, authors' characteristics were extracted and compared. One of the most interesting features is that authority and susceptibility are not absolute psychological characteristics of authors,
but depend on the contest they work. The majority of authors, even when they are leaders in one field (and hence strongly influence that community), do not exhibit the same authority in the other. In other words one can be source of new ideas in a sector while being strongly receptive in an other.  Generally speaking the APS community is more coesive and authors tend to share collaborators' interests more than those working in computer science. 

It is our opinion that general results are robust against the model approximation, however it worth stating some of the most relevant: 

The \textbf{aging of the links} \cite{zhu2003effect} \cite{tutoky2013weights}. In this model, once a link is established it is supposed to hold forever; whereas, in reality, links can also vanish for several reasons related to competitiveness, displacements, personal frictions etc. In general, here the strengths of links and their evolution are not taken into account. 
 
The \textbf{multiplex problem} \cite{mucha2010community,gomez2013diffusion,granell2013dynamical}. This paper deals with two interacting social layer only (based on co-authorship), however, people have different 
types of relationships (such as for instance working in the same company; participating same meetings etc) and, hence, other layers should be accounted for the interest diffusion. 
 
The \textbf{semantic profiling} \cite{middleton2004ontological,sheth2001system}. Generally speaking, one is not allowed to assume that there exist a set of disjoint topics covering all possible interests. Concepts are normally overlapping and, possibly, one interest may induce another one in a "close" topic. In the present model, in order to provide members with a  semantic profile,  the existence of a basic set of disjoint topics is strictly required. Further developing will enter the semantic structure of the domain of interest and will lead to more complex modelling.

\textbf{Psychological types}. The present model only allows people to be influenced by friends (or by the environment) or to be independent on them. 
However, there are several reasons for which a person may deliberately decide to do things, not just disregarding friends' positions, but in contrast with them. 
This may take place for competition or just for spirit of independence. This type of behaviour is usually referred to as "anti conformity"  and it has been studied elsewhere \cite{nyczka2013anticonformity,willis1965conformity,pronin2007alone}.

To conclude, one may state that the preliminary analysis performed on the  APS and DBLP datasets demonstrates that diffusion of interest on social networks is a reality and historical data can be analysed to provide information on 
members' profiles and their human relationships. The Computer Science and the APS communities form a Semantic Multiplex. The analysis on their common authors provides insights on the similarity and differences of the two scientific communities.

\section{Acknowledgements}
We acknowledge interesting discussions with Antonio Scala, Walter Quattrociocchi and Alberto Tofani and the American Physical Society for providing us the dataset.




\end{document}